\documentclass[aps,twocolumn,showpacs,superscriptaddress]{revtex4}

\usepackage{graphicx}
\usepackage{bm}
\usepackage{amsmath}
\usepackage{dcolumn}%

\newcommand{\dxy}{\ensuremath{d_{xy}}}
\newcommand{\dxz}{\ensuremath{d_{xz}}}
\newcommand{\dyz}{\ensuremath{d_{yz}}}
\newcommand{\dzz}{\ensuremath{d_{3z^2-1}}}
\newcommand{\dxxyy}{\ensuremath{d_{x^2-y^2}}}

\newcolumntype{/}{D{/}{/}{2,2}}  
\newcolumntype{.}{D{.}{.}{0}}  

\begin{document}

\title{Electronic structure and resonant inelastic x-ray scattering in
  Ta$_2$NiSe$_5$ }

\author{D.A. Kukusta}

\affiliation{G. V. Kurdyumov Institute for Metal Physics of the
  N.A.S. of Ukraine, 36 Academician Vernadsky Boulevard, UA-03142
  Kyiv, Ukraine}

\author{L.V. Bekenov}

\affiliation{G. V. Kurdyumov Institute for Metal Physics of the
  N.A.S. of Ukraine, 36 Academician Vernadsky Boulevard, UA-03142
  Kyiv, Ukraine}

\author{A.N. Yaresko}

\affiliation{Max-Planck-Institute for Solid State Research,
  Heisenbergstrasse 1, 70569 Stuttgart, Germany}

\author{K. Ishii}

\affiliation{Synchrotron Radiation Research Center, National
  Institutes for Quantum Science and Technology, Sayo, Hyogo 679-5148,
  Japan }
 
\author{T. Takayama}

\affiliation{Max-Planck-Institute for Solid State Research,
  Heisenbergstrasse 1, 70569 Stuttgart, Germany}

\author{H. Takagi}

\affiliation{Max-Planck-Institute for Solid State Research,
  Heisenbergstrasse 1, 70569 Stuttgart, Germany}

\author{V.N. Antonov}

\affiliation{G. V. Kurdyumov Institute for Metal Physics of the
  N.A.S. of Ukraine, 36 Academician Vernadsky Boulevard, UA-03142
  Kyiv, Ukraine}

\affiliation{Max-Planck-Institute for Solid State Research,
  Heisenbergstrasse 1, 70569 Stuttgart, Germany}

\date{\today}

\begin{abstract}

  We study the electronic structure of Ta$_2$NiSe$_5$ in its
  low-temperature semiconducting phase, using resonant inelastic x-ray
  scattering (RIXS) at the Ta $L_3$ edge. We also investigate the
  electronic properties of Ta$_2$NiSe$_5$ within the
  density-functional theory (DFT) using the generalized gradient
  approximation in the framework of the fully relativistic
  spin-polarized Dirac linear muffin-tin orbital band-structure
  method. While ARPES, dc transport, and optical measurements indicate
  that Ta$_2$NiSe$_5$ is a small band-gap semiconductor, DFT gives a
  metallic nonmagnetic solution in Ta$_2$NiSe$_5$. To obtain the
  semiconducting ground state in Ta$_2$NiSe$_5$ we use a
  self-interaction-like correction procedure by introducing an
  orbital-dependent potential $V_l$ into the Hamiltonian. We
  investigate theoretically the x-ray absorption spectroscopy (XAS)
  and RIXS spectra at the Ni and Ta $L_3$ edges and analyze the
  spectra in terms of interband transitions. We also investigate the
  RIXS spectra as a function of momentum transfer vector {\bf Q} and
  incident photon energy.
  
\end{abstract}


\maketitle

\section{Introduction}

\label{sec:introd}

Transition-metal compounds containing 5$d$ elements usually have a rich
variety of novel electronic states emerged from competing interactions,
including the on-site Coulomb repulsion $U$, crystal-electric field (CEF),
and spin-orbit coupling (SOC). These compounds possess the on-site Coulomb
repulsion and SOC of the same order of magnitude ($U$ $\sim$ 1-2 eV,
$\lambda_{SOC}$ $\sim$ 0.5 eV) \cite{WCK+14}. It can give rise to some
fascinating phenomena, such as topological insulators
\cite{QZ10,Ando13,WBB14,BLD16}, Mott insulators
\cite{KJM+08,KOK+09,JaKh09,WSY10,MAV+11}, Weyl semimetals
\cite{WiKi12,GWJ12,SHJ+15}, and quantum spin liquids \cite{JaKh09,KAV14}.

In this paper, we consider the transition-metal chalcogenide
Ta$_2$NiSe$_5$, which has been studied recently in the respect of the
possible realization of the excitonic insulator (EI) state
\cite{WST+09,WST+12,KTK+13}. This chalcogenide form a
quasi-one-dimensional (1D) chain structure with Ni single and Ta
double chains running along the $a$ axis. Along the $b$ axis the
layers are loosely held together by van der Waals' forces
\cite{SuIb85}. High temperature measurements of resistivity have shown
that Ta$_2$NiSe$_5$ is a small band gap semiconductor with quasi-1D
anisotropic electron conduction \cite{SCF+86}. The nature of the
electronic structure of the Ta$_2$NiSe$_5$ orthorhombic phase (space
group $Cmcm$) is still strongly debated and this phase is argued to be
a semimetal \cite{FSJ+19,WMM+20,LKE+19}, almost zero energy gap
semiconductor with the energy gap $\Delta$E $\le$ 0.05 eV
\cite{LKL+17}, or conventional semiconductor
\cite{WST+12,SWK+14,MHG+17}. At critical temperature $T_c$ = 328 K,
however, an anomaly in the resistivity occurs, which was associated
with the second-order structural phase transition from an orthorhombic
to monoclinic phase \cite{SCF+86,KTK+13}. The resistivity exhibits
insulating behavior both above and below the transition temperature
$T_c$ \cite{SWK+14}. The magnetic susceptibility exhibits diamagnetism
in a wide temperature range (4.2-900 K) and shows a sudden drop (being
more negative) $\le T_c$ \cite{SCF+86}. The characteristic energy gap
according to optical conductivity and RIXS measurements reaches
$\Delta$E $\sim$0.16 eV $\le T_c$
\cite{LKL+17,LYP+17,LRK+21}. Tunneling spectroscopy estimates the
energy gap $\sim$0.3 eV at 78 K \cite{LKE+19}. ARPES experiments
showed that the spectra are strongly temperature dependent
\cite{WST+09,WST+12}.  At 40 K the valence band top flattens, the
quasiparticle peak sharpens and the size of the band gap becomes
larger. It was suggested that the EI state is realized as the ground
state of this material, with excitons formed by a charge transfer
between the Ni and Ta chains \cite{WST+09,SWK+14,WTH+18}. However,
direct evidence of excitons and their behavior across the EI
transition has thus far not been reported.

While ARPES, dc transport, and optical measurements indicate that
Ta$_2$NiSe$_5$ is a small band-gap semiconductor, density functional
theory (DFT) gives a metallic nonmagnetic solution in Ta$_2$NiSe$_5$
\cite{KTK+13,LYP+17}. The top of the valence band, formed by (almost)
completely occupied Se 4$p$ and Ni 3$d$ states, overlaps with the
bottom of the conduction band, which is derived from Ta 5$d$ $t_{2g}$
states hybridized with the chalcogen 4$p$ ones. Thus, according to the
DFT calculations the formal valencies are close to Ta$^{5+}$
(5$d^{0}$), Ni$^{0}$ (3$d^{10}$), and Se$^{2-}$ (4$p^{6}$). Because
Ta$_2$NiSe$_5$ possesses only fully occupied (Ni 3$d$ and Se 4$p$) and
completely empty (Ta 5$d$) shells it cannot be the subject for the
application of the LDA+$U$ method. This method is not able to open up
a gap \cite{LYP+17,LKE+19}. To reproduce the experimental band gap
Kaneko {\it et al.} \cite{KTK+13} introduce a self-interaction-like
correction (SIC) procedure \cite{PZ81,LMW+10}, where the conduction
(valence) bands are shifted upward (downwards) by adding a SIC-like
orbital-dependent potential $V_l$ to the Hamiltonian so as to open up
the gap in the band dispersion. The authors add the potentials +$V_l$
to the energy of the 5$d$ orbitals of Ta and $-V_l$ to the energy of
the 3$d$ orbitals of Ni and 4$p$ orbitals of Se in the orthorhombic
structure. They found that the band gap actually opens up for $V_l$ =
4.2 eV. They used the value of $V_l$ = 5.0 eV in their band structure
calculations. A similar procedure was used also in Ref. \cite{LYP+17}
to calculate the electronic structure and optical spectra in
monoclinic Ta$_2$NiSe$_5$.  They found that the best agreement between
the theoretically calculated and experimentally measured optical
spectra is obtained when applying the orbital-dependent potential
$V_l$ only to the Se 4$p$ states ($V_{Se_{4p}}$ = $-$2.5 eV).

Ma {\it et al.} \cite{MWM+22} investigate the electronic structure of
Ta$_2$NiSe$_5$ as a promising candidate to a three-dimensional
topological excitonic insulator. They obtained the Dirac cone type
surface states of the low-temperature monoclinic phase.

Windg\"atter {\it et al.} \cite{WRM+21} provide extensive
first-principles calculations of the electronic structure of
Ta$_2$NiSe$_5$ and Ta$_2$NiS$_5$ using the DFT as well as G$_0$W$_0$
approximations. Authors found very strong gap dependence on the type
of exchange correlation potential, the gape is equal to 0.018, 0.120,
0.179, and 0.278 eV for for the Perdew-Burke-Ernzerhof (PBE), mBJ,
HSE03, and HSE06 potentials, respectively, with taking into account
SOC. In all cases, G$_0$W$_0$ approach predicts a bandgap between 0.1
and 0.163 eV, which is in good agreement with the experimental gap of
0.16 eV measured in optics \cite{LKL+17,LYP+17,LRK+21}. The
temperature dependence of the band structure shows that the bandgap
never closes for increasing electronic temperatures but shows the
expected renormalization effects of standard semiconductors due to the
increased carrier density. They also calculate the phonon dispersion
as well as electron-phonon interaction in the compounds.

In the present study, we focus on the electronic structure and
resonant inelastic x-ray spectra (RIXS) of Ta$_2$NiSe$_5$. RIXS
measurements have been successfully performed at the Ni $L_3$ edge for
Ta$_2$NiSe$_5$ by Monney {\it et al.}  \cite{MHP+20}. The authors
present a RIXS map as a function of the incident photon energy,
measured at the Ni $L_3$ edge as well as the in-plane momentum $Q_x$
dispersion with $\sigma$-polarized incident light at 30 K. They also
present the x-ray absorption spectroscopy (XAS) spectrum measured at
the Ni $L_3$ edge by the total fluorescence yield. They obtained a
band gap at the center of the Brillouin zone of $\sim$0.35 eV at
$\le T_c$. The authors also estimated the effective valence $m_V$ and
conductive $m_C$ band masses in close vicinity to the Fermi
level. These masses were found to be rather large: $m_V$ = 0.8$m_e$
and $m_C$ $\sim$ 0.9$\div$1.3$m_e$. Lu {\it et al.} also measured the
RIXS spectra at the Ta and Ni $L_3$ edges in a small energy interval
up to 1.2 eV \cite{LRK+21}. Below $T_c$, their RIXS energy-momentum
map shows a band gap at the Brillouin zone center of $\sim$0.16
eV. The authors interpreted the RIXS spectra in terms of the
momentum-resolved joint density of states (JDOS) without taking into
account corresponding matrix elements.

Although there has been great progress in the RIXS experiments over the past
decade, the number of theoretical calculations of RIXS spectra is
extremely limited. The most calculations of the RIXS spectra of various
materials have been carried out using the atomic multiplet approach with several
adjustable parameters. The question is to what extent DFT is able to reveal the
aspects of RIXS spectra. In this paper, we report the experimentally measured
RIXS spectrum of Ta$_2$NiSe$_5$ at the Ta $L_3$ edge in a wide energy interval
as well as a detailed theoretical investigation of the electronic structure
and RIXS spectra at the Ta and Ni $L_3$ edges. The energy band structure of
this transition metal chalcogenide was calculated using the fully relativistic
spin-polarized Dirac linear muffin-tin orbital band-structure method.

The paper is organized as follows. The crystal structure of Ta$_2$NiSe$_5$ and
computational details are presented in Sec. II. Section III presents the
electronic structure of the chalcogenide. In Sec. IV theoretical
investigations of the RIXS spectra of Ta$_2$NiSe$_5$ at the Ni and Ta $L_3$
edges are presented, and the theoretical results are compared with
experimental measurements. Here, we also report the results of the
calculations of x-ray absorption spectra at the Ni and Ta $L_3$
edges. Finally, the results are summarized in Sec. V.

\section{Experimental and computational details}
\label{sec:details}

\subsection{RIXS} 

Resonant inelastic x-ray scattering refers to the process where the
material first absorbs a photon. The system then is excited to a
short-lived intermediate state, from which it relaxes radiatively. In
an experiment, one studies the x rays emitted in this decay
process. In the direct RIXS process \cite{AVD+11} an incoming photon
with energy $\hbar \omega_{\mathbf{k}}$, momentum $\hbar \mathbf{k}$
and polarization $\bm{\epsilon}$ excites the solid from a ground state
$|{\rm g}\rangle$ with energy $E_{\rm g}$ to the intermediate state
$|{\rm I}\rangle$ with energy $E_{\rm I}$. During relaxation the
outcoming photon with energy $\hbar \omega_{\mathbf{k}'}$, momentum
$\hbar \mathbf{k}'$ and polarization $\bm{\epsilon}'$ is emitted, and
the solid is in the state $|{\rm f}\rangle$ with energy $E_{\rm f}$. A
valence electron is excited from state $\mathbf{k}$ to states
$\mathbf{k}'$ with energy $\hbar \omega = \hbar \omega_{\mathbf{k}} -
\hbar \omega_{\mathbf{k}'}$ and momentum transfer $\hbar \mathbf{q}$ =
$\hbar \mathbf{k} - \hbar \mathbf{k}'$.
The RIXS intensity can in general be presented in terms of a
scattering amplitude as \cite{AVD+11}

\begin{eqnarray}
I(\omega, \mathbf{k}, \mathbf{k}', \bm{\epsilon}, \bm{\epsilon}')
&=&\sum_f \left| T_{fg}(\mathbf{k}, \mathbf{k}',
\bm{\epsilon}, \bm{\epsilon}', \omega_{\mathbf{k}}) \right|^2 \nonumber \\
&&\times \delta(E_f+\hbar \omega_{\mathbf{k}'}-E_g-\hbar \omega_{\mathbf{k}})\, ,
\label{I}
\end{eqnarray}
where the delta function enforces energy conservation and the
amplitude $T_{fg}(\mathbf{k}, \mathbf{k}', \bm{\epsilon},
\bm{\epsilon}', \omega_{\mathbf{k}})$ reflects which excitations are
probed and how, for instance, the spectral weights of final state
excitations depend on the polarization vectors $\bm{\epsilon}$ and
$\bm{\epsilon'}$ of the incoming and outgoing x-rays,
respectively.

Our implementation of the code for the calculation of the RIXS
intensity uses Dirac four-component basis functions \cite{NKA+83} in
the perturbative approach \cite{ASG97}. RIXS is a second-order
process, and its intensity is given by

\begin{eqnarray}
I(\omega, \mathbf{k}, \mathbf{k}', \bm{\epsilon}, \bm{\epsilon}')
&\propto&\sum_{\rm f}\left| \sum_{\rm I}{\langle{\rm
    f}|\hat{H}'_{\mathbf{k}'\bm{\epsilon}'}|{\rm I}\rangle \langle{\rm
    I}|\hat{H}'_{\mathbf{k}\bm{\epsilon}}|{\rm g}\rangle\over
  E_{\rm g}-E_{\rm I}} \right|^2 \nonumber \\ && \times
\delta(E_{\rm f}-E_{\rm g}-\hbar\omega),
\label{I1}
\end{eqnarray}
where the delta function enforces energy conservation, and the photon
absorption operator in the dipole approximation is given by the
lattice sum
$\hat{H}'_{\mathbf{k}\bm{\epsilon}}=
\sum_\mathbf{R}\hat{\bm{\alpha}}\bm{\epsilon} \exp(-{\rm
  i}\mathbf{k}\mathbf{R})$, where $\hat{\bm{\alpha}}$ are Dirac
matrices. Both $|{\rm g}\rangle$ and $|{\rm f}\rangle$ states are
dispersive so the sum over final states is calculated using the linear
tetrahedron method \cite{LeTa72}. The matrix elements of the RIXS
process in the frame of the fully relativistic Dirac LMTO method were
presented in our previous publications \cite{unpub:KuYar19,AKB22a}.

\subsection{Crystal structure} 

Ta$_2$NiSe$_5$ at high temperature possesses an orthorhombic crystal
structure (space group $Cmcm$, number 63). The material has a layered
structure stacked loosely by a weak van der Waals interaction, and in
each layer, Ni single chains and Ta double chains are running along
the $a$ axis of the lattice to form a quasi-one-dimensional (1D) chain
structure \cite{SuIb85}. At critical temperature $T_c$ = 326 K,
Ta$_2$NiSe$_5$ undergoes a second-order transition from an orthorhombic
structure to a structure with a subtle monoclinic distortion and reduced
electrical conductivity \cite{SuIb85,SCF+86}.

\begin{figure}[tbp!]
\begin{center}
\includegraphics[width=1.\columnwidth]{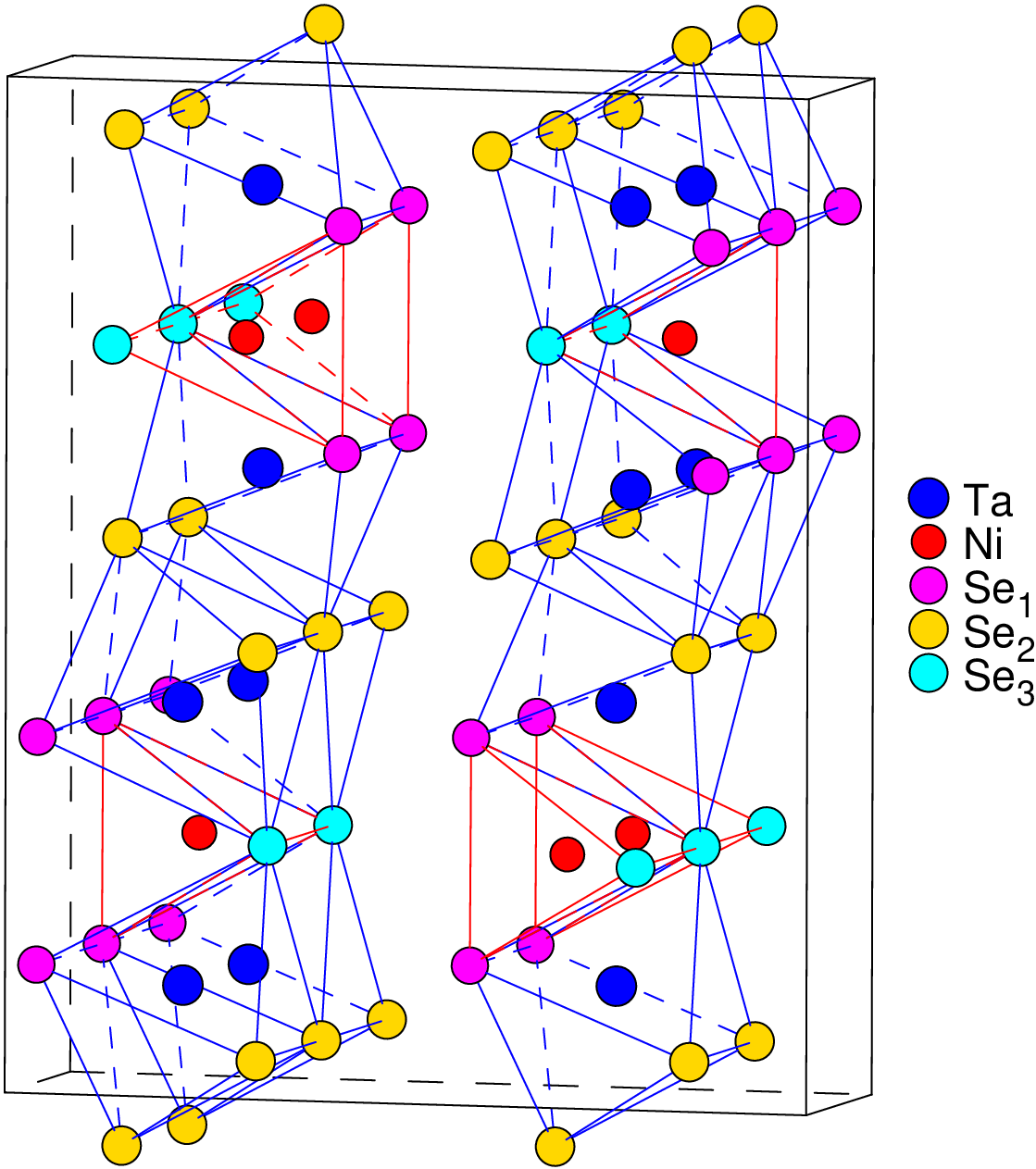}
\end{center}
\caption{\label{struc_TNS}(Color online) The monoclinic crystal structure of
  Ta$_2$NiSe$_5$ (space group $C12/c1$, number 15) \cite{SuIb85}. Blue and
  red spheres represent Ta and Ni atoms, respectively, magenta, yellow, and
  green spheres show Se atoms. }
\end{figure}

Figure \ref{struc_TNS} shows the monoclinic crystal structure of
Ta$_2$NiSe$_5$ (space group $C12/c1$, number 15) \cite{SuIb85}. The Ni and Ta
atoms are tetrahedrally and octahedrally coordinated by chalcogen atoms,
respectively. The polyhedra are combined by sharing edges to form a repeated
unit Ta-Ni-Ta. These chains, which stack in the $a$ direction, are connected
via edge-shared octahedra and shared vertices of tetrahedra
(Fig. \ref{struc_TNS}). The unit cell contains two such chains
displaced by $a$/2 with respect to each other. The coordination about the Ta
and Ni atoms is slightly distorted from ideal geometry (Table
\ref{struc_tab_TNS}).

\begin{table}[tbp!]
  \caption {The atomic positions of Ta$_2$NiSe$_5$. The lattice constants are
    equal to $a$ = 3.4974 \AA, $b$ = 12.8460 \AA\, and $c$ = 15.6457
    \AA, $\beta$ = 90.0$^{\circ}$) for the orthorhombic crystal
    structure $Cmcm$ \cite{WMM+20} and $a$ = 3.496 \AA, $b$ = 12.829
    \AA\, and $c$ = 15.641 \AA, $\beta$ = 90.530$^{\circ}$ for the
    monoclinic structure $C12/c1$ \cite{SuIb85}. }
\label{struc_tab_TNS}
\begin{center}
\begin{tabular}{|c|c|c|c|c|}
\hline
Structure(Ref.)    & Atom     & $x$         & $y$      & $z$     \\
\hline
                     & Ta     &  0.          & 0.22123  &  0.11025  \\
                     & Ni     &  0.          & 0.70104  &  0.25   \\
$Cmcm$ \cite{WMM+20} & Se$_1$ &  0.5         & 0.08037  &  0.13776  \\
                     & Se$_2$ &  0.          & 0.14583  &  0.95074   \\
                     & Se$_3$ &  0.          & 0.32710  &  0.25    \\
\hline
                      & Ta     &  $-$0.00793 & 0.221349 &  0.110442  \\
                      & Ni     &  0.         & 0.70113  &  0.25   \\
$C12/c1$ \cite{SuIb85}& Se$_1$ &  0.50530    & 0.080385 &  0.137979  \\
                      & Se$_2$ &  $-$0.00513 & 0.145648 &  0.950866   \\
                      & Se$_3$ &  0.         & 0.32714  &  0.25    \\
\hline
\end{tabular}
\end{center}
\end{table}

The primitive unit cell contains four Ta ions, two Ni ions, and ten Se
ions. All Ta and Ni sites are crystallographically equivalent but
there are three inequivalent Se sites, Se$_1$, Se$_2$ and Se$_3$. The
Ta-Ta inter-chain distances are fairly large $d^{\perp}_{\rm{Ta-Ta}}$
= 3.903, 3.980 \AA\,, so that there is no significant Ta-Ta
bonding. Around each Ni atom, however, there are four Ta atoms in a
square-planar arrangement with very short Ta-Ni distances
$d_{\rm{Ta-Ni}}$ = 2.804, 2.813 \AA. The Ni ions have a tetrahedra
NiSe$_4$ arrangement with $d_{\rm{Ni-Se_1}}$ = 2$\times$2.339 \AA,
$d_{\rm{Ni-Se_3}}$ = 2$\times$2.381 \AA. The Ta$^{5+}$ cations are
surrounded by Se octahedrons with $d_{\rm{Ta-Se_1}}$ = 2.523, 2.581
\AA\, $d_{\rm{Ta-Se_2}}$ = 2.588, 2.661, 2.678 \AA\,, and
$d_{\rm{Ta-Se_3}}$ = 2.570 \AA\, interatomic distances. The distance
between metals chained along $x$: $d^{\parallel}_{\rm{Ta-Ta}}$ =
$d^{\parallel}_{\rm{Ni-Ni}}$ = 2$\times$3.496 \AA\, \cite{SuIb85}.

\subsection{Experimental details} 

The resonant inelastic x-ray scattering measurement on the Ta
$L_3$-edge was performed at BL11XU SPring-8. The incident x-ray was
monochromatized by a double-crystal Si(111) monochromator and by a
secondary 4-bounce Si(333) asymmetric monochromator. $\pi$-polarized
x-rays with the energy of 9.875 keV were irradiated onto the ac-plane
of the Ta$_2$NiSe$_5$ crystal, and the horizontally-scattered x-rays
were energy-analyzed by a Ge(840) diced analyze and collected by the
Mythen microstrip x-ray detector (Dectris). The total energy
resolution was 90 meV. The measurement was performed at a low
temperature $\sim$10 K.

\subsection{Calculation details}

The details of the computational method are described in our previous
papers \cite{AJY+06,AHY+07b,AYJ10,AKB22a} and here we only mention
several aspects. The band structure calculations were performed using
the fully relativistic LMTO method \cite{And75,book:AHY04}. This
implementation of the LMTO method uses four-component basis functions
constructed by solving the Dirac equation inside an atomic sphere
\cite{NKA+83}. The exchange-correlation functional of the GGA-type was
used in the version of Perdew, Burke and Ernzerhof \cite{PBE96}. The
Brillouin zone integration was performed using the improved
tetrahedron method \cite{BJA94}. The basis consisted of Ta and Ni $s$,
$p$, $d$, and $f$; and Se $s$, $p$, and $d$ LMTO's.

It is widely believed that the $d-d$ excitations show only small
momentum transfer vector {\bf Q} dependence in 5$d$ transition metal
compounds \cite{LKH+12,KTD+20}. However, the soft RIXS spectra in 3$d$
transition metals are more sensitive to the value of {\bf Q}.  We used
in our RIXS calculations {\bf Q} = (0.0625, 1.25, 0) at the Ni $L_3$
and {\bf Q} = (0, 14.5, 0) at the Ta $L_3$ edge, which have been used
in the corresponding experimental measurements \cite{MHP+20}. We also
investigate the dispersion of the RIXS spectra at the Ni and Ta $L_3$
edges as a function of $Q_x$.

The finite lifetime of a core hole was accounted for by folding the spectra
with a Lorentzian. The widths of core levels $\Gamma$ for Ta, Ni, and Se were
taken from Ref. \cite{CaPa01}. The finite experimental resolution of the
spectrometer was accounted for by a Gaussian of 0.6 eV (the $s$ coefficient of
the Gaussian function).

Note that in our electronic structure calculations, we rely on experimentally
measured atomic positions and lattice parameters \cite{SuIb85,WMM+20} (see
Table \ref{struc_tab_TNS}) because they are well established for these
materials and are probably still more accurate than those obtained from DFT.

It is known that DFT band calculations usually underestimate the band gap in
semiconductors \cite{book:AHY04}. Because Ta$_2$NiSe$_5$ possesses only fully
occupied (Ni 3$d$ and Se 4$p$) and completely empty (Ta 5$d$) shells the
LDA+$U$ method is not able to open up a gap \cite{LYP+17,LKE+19}. To reproduce
a semiconducting ground state in Ta$_2$NiSe$_5$ we used a
self-interaction-like correction procedure as it was proposed by Kaneko {\it
  et al.}  \cite{KTK+13}, where the conduction (valence) bands are shifted
upward (downwards) by adding a SIC-like orbital-dependent potential $V_l$ to
the Hamiltonian. We used $V_l$ as a parameter an adjusted it to produce the
correct value of the band gap and the best agreement with the RIXS
experimental spectra.

\section{Electronic structure}
\label{sec:bands}

We found that Ta$_2$NiSe$_5$ is a direct-gap semiconductor with the
gap minimum at the $\Gamma$ point of the Brillouin zone in agreement
with the experiment \cite{WST+09,SCF+86}. The energy gap for the low
temperature $C12/c1$ phase of $\sim$0.16 eV was estimated from optical
measurements \cite{LKL+17,LYP+17,LRK+21}. Table \ref{Eg_TNS} presents
the theoretically calculated energy gap $\Delta E$ (in eV) for
different $V_l$ for the low temperature $C12/c1$ phase.

\begin{table}[tbp!]
  \caption{\label{Eg_TNS} The theoretically calculated energy gap
    $\Delta E$ (in eV) for different $V_l$. The experimental energy
    gap for the low temperature $C12/c1$ phase was estimated to be
    $\sim$0.16 eV \cite{LKL+17,LYP+17,LRK+21}. }
\begin{center}
\begin{tabular}{cccccccc}
\hline
$V_{Ta_{5d}}$ & $V_{Ni_{3d}}$ & $V_{Se_{4p}}$ & $\Delta E$  \\
\hline
 0.0   &  0.0    &  -2.5 &  0.0790  \\
 0.0   &  0.0    &  -4.0 &  0.1430  \\
 0.0   &  0.0    &  -5.0 &  0.1634  \\
 0.0   & -2.5    &  -2.5 &  0.0964  \\
 0.0   & -4.0    &  -2.5 &  0.1123  \\
 0.0   & -5.0    &  -2.5 &  0.1125  \\
 0.0   & -2.5    &  -5.0 &  0.2016  \\
 0.0   & -4.0    &  -5.0 &  0.2282  \\
 0.0   & -5.0    &  -5.0 &  0.2468  \\
 2.5   &  0.0    &  -2.5 &  $<$0    \\
 2.5   & -2.5    &   0.0 &  $<$0    \\
 2.5   & -2.5    &  -2.5 &  $<$0    \\
 5.0   &  0.0    &  -5.0 &  0.4791  \\
 5.0   & -5.0    &  -5.0 &  0.7907  \\
\hline
\end{tabular}
\end{center}
\end{table}

The shift of Se 4$p$ states is the most important in this case. We
found that the best agreement for the band gap in this phase can be
achieved for $V_{Se_{4p}}$ = $-$5.0 eV with zero values for the other
two parameters (see Table \ref{Eg_TNS}). It should be mentioned that
the self-interaction is the interaction of an electron with its own
negative density \cite{PZ81,LMW+10}. Thus, it shifts one electron
state to higher energies. Therefore, the use of the negative potential
shift for the Se 4$p$ and Ni 3$d$ states in order to simulate the
self-interaction correction is correct. But the addition of a positive
potential shift to the Ta 5$d$ states cannot be justified in the frame
of the SIC-like method. Moreover, it worsens the agreement of
calculated optical spectra \cite{LYP+17} as well as the RIXS spectra
(see Fig. \ref{rixs_Ni_Vl} below) with experimental measurements. It
also cannot produce the correct energy band gap (Table \ref{Eg_TNS}).

\begin{figure}[tbp!]
\begin{center}
\includegraphics[width=0.99\columnwidth]{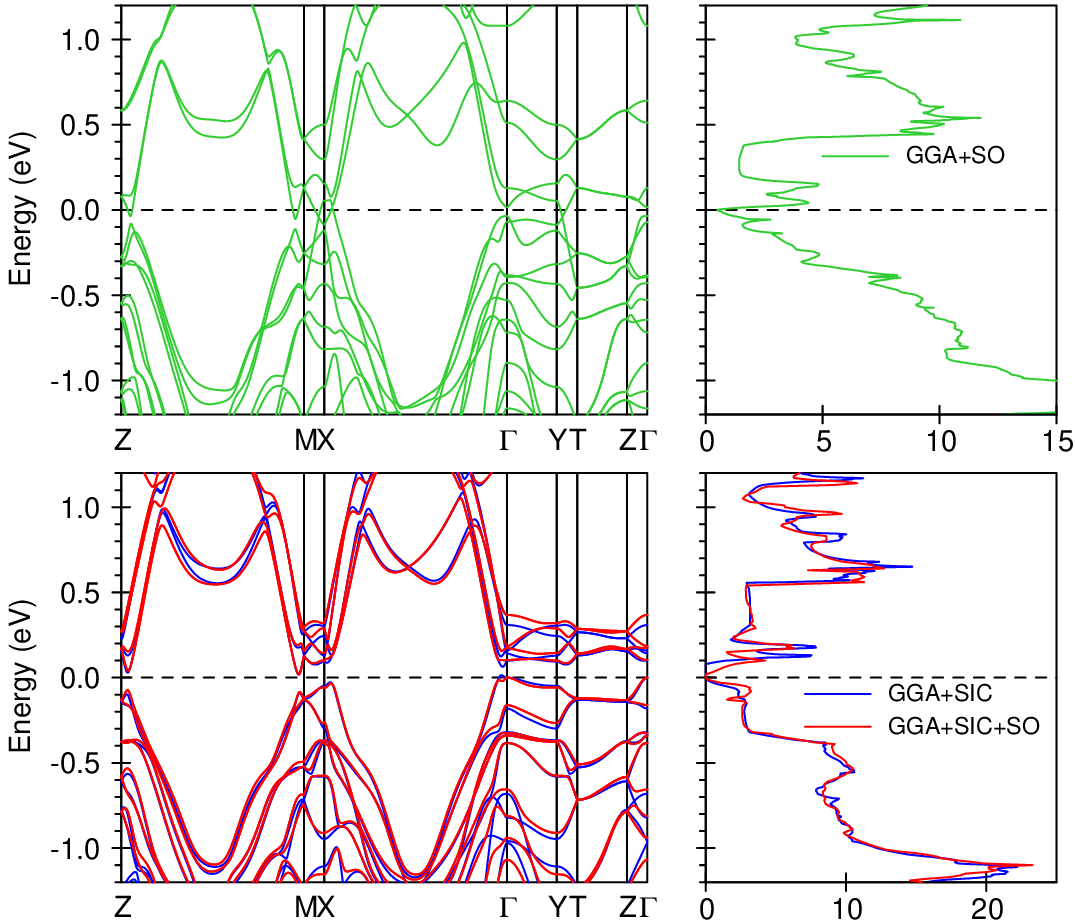}
\end{center}
\caption{\label{BND_TNO_SO}(Color online) The energy band structure
  and total density of states (DOS) [in states/(atom eV)] of
  monoclinic Ta$_2$NiSe$_5$ calculated in the fully relativistic Dirac
  GGA+SO approximation (the upper panel, colors represent the dominant
  orbital character as marked in the legend) and with taking into
  account the self-interaction-like correction (SIC) ($V_{Se_{4p}}$ =
  $-$5.0 eV) with (red curves) and without (blue curves) SOC (the
  lower panel). }
\end{figure}

Figure \ref{BND_TNO_SO} presents the energy band structure and total
DOS of monoclinic Ta$_2$NiSe$_5$ calculated in the fully relativistic
Dirac GGA+SO approximation (the top panel) and with taking into
account the SIC-like correction ($V_{Se_{4p}}$ = $-$5.0 eV) with (red
curves) and without (blue curves) SOC (the lower panel). We found that
SOC plays a minor role in the electronic structure and band gap value
in Ta$_2$NiSe$_5$.

\begin{figure}[tbp!]
\begin{center}
\includegraphics[width=0.99\columnwidth]{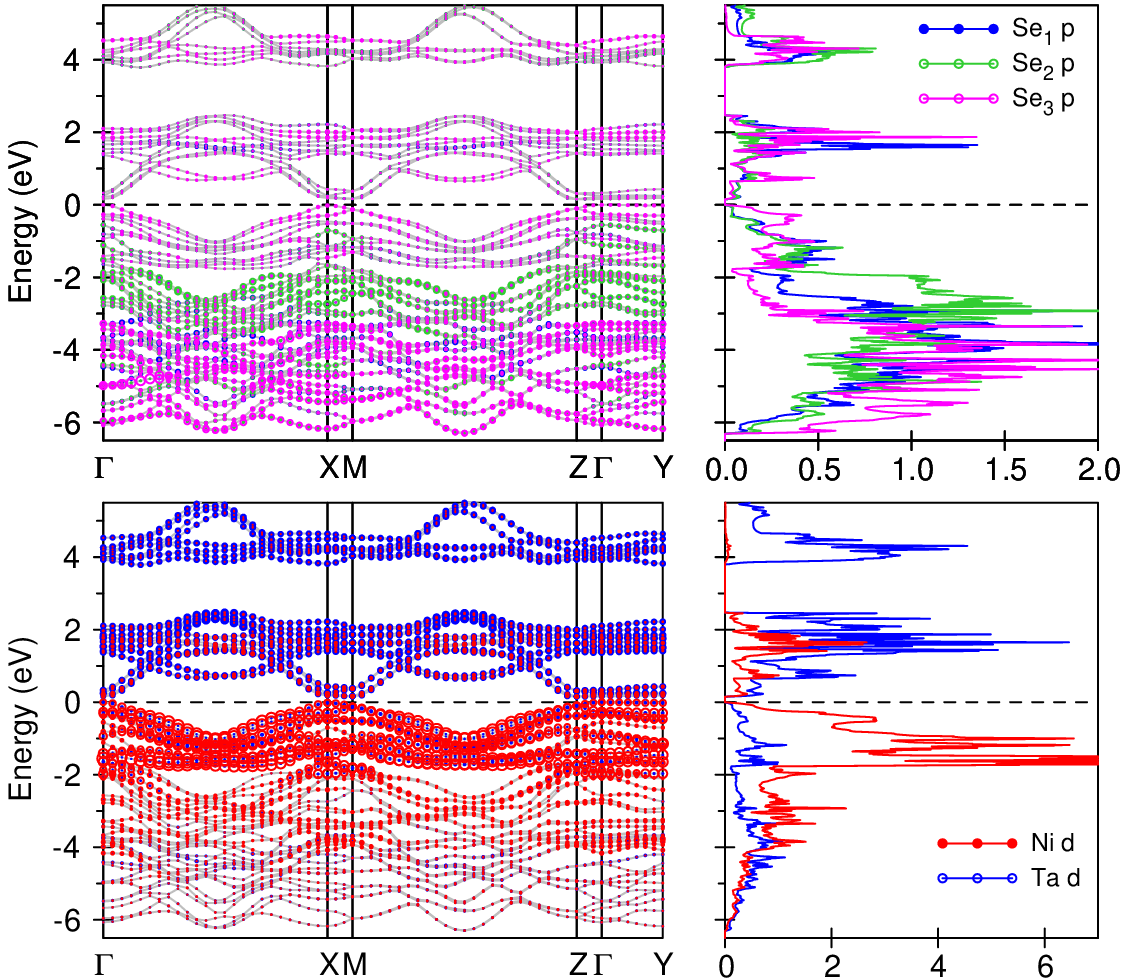}
\end{center}
\caption{\label{BND_TNS_FB}(Color online) The energy band structure
  and partial density of states (DOS) of monoclinic Ta$_2$NiSe$_5$
  calculated in the fully relativistic Dirac GGA+SIC+SO approximation
  ($V_{Se_{4p}}$ = $-$5.0 eV). Colors represent the dominant orbital
  character as marked in the legend. }
\end{figure}

\begin{figure}[tbp!]
\begin{center}
\includegraphics[width=0.99\columnwidth]{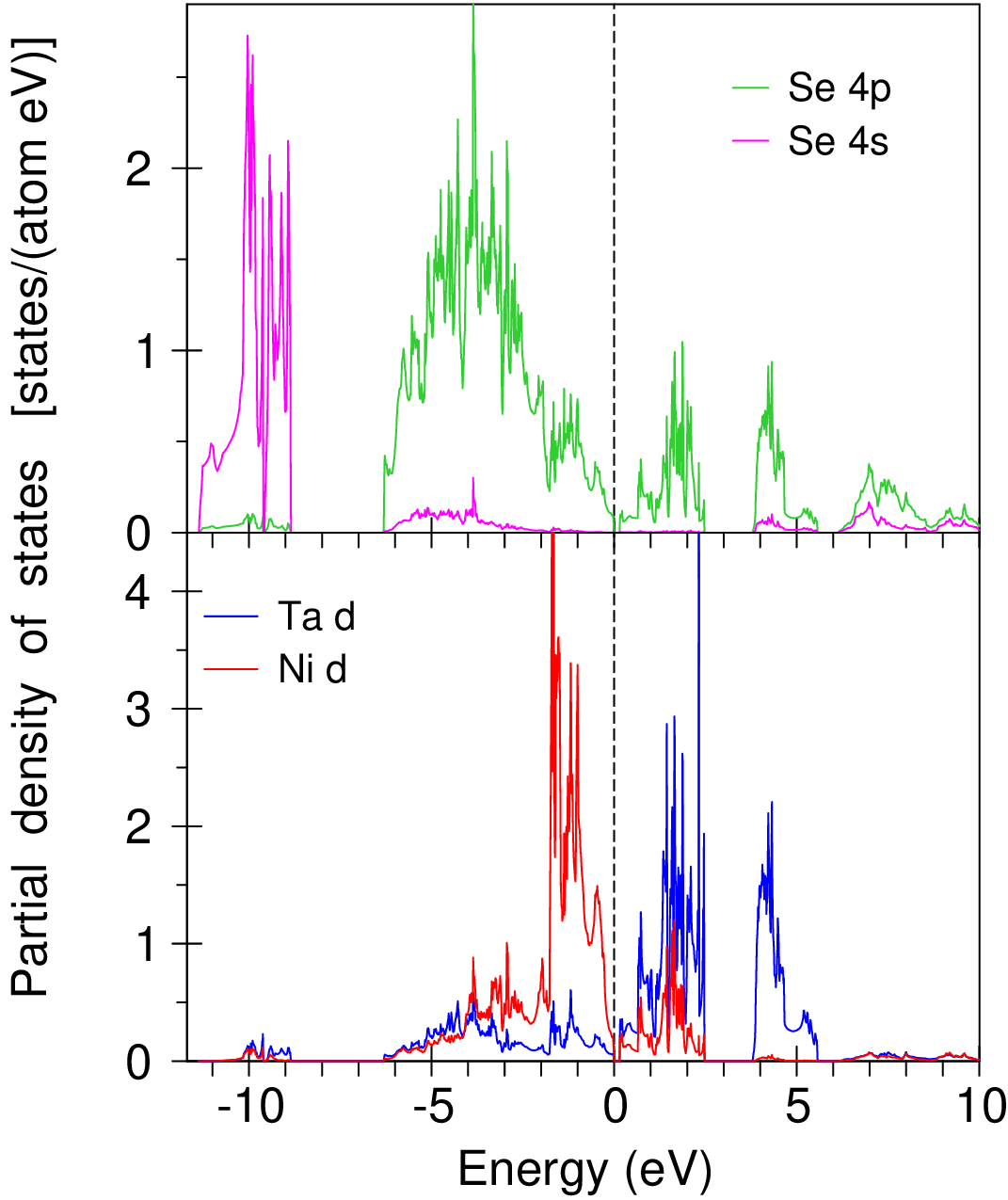}
\end{center}
\caption{\label{PDOS_TNS}(Color online) The partial density of states (DOS) of
  monoclinic Ta$_2$NiSe$_5$ calculated in the fully relativistic Dirac
  GGA+SIC+SO approximation ($V_{Se_{4p}}$ = $-$5.0 eV). }
\end{figure}

Figures \ref{BND_TNS_FB} and \ref{PDOS_TNS} show the energy band structure and
partial density of states (PDOS) of monoclinic Ta$_2$NiSe$_5$ calculated in
the fully relativistic Dirac GGA+SIC+SO approximation ($V_{Se_{4p}}$ = $-$5.0
eV). Se 4$s$ states are located from $-$11.4 to $-$8.85 eV, Se 4$p$ states
occupy a rather large energy interval from $-$6.3 to 2.5 eV and between 3.8
and 10 eV. They are strongly hybridize with the Ni 3$d$ valent states and Ta
5$d$ conduction states. The fully occupied Ni 3$d$ states are situated between
$-$1.9 eV and $E_F$. There are some Ni 3$d$ states at the bottom of the valent
band between $-$6.3 and $-$1.9 eV, which occur from the strong hybridization
with Se 4$p$ states. Ta {\ensuremath{t_{2}}} states are situated between 0.16
and 2.5 eV. Ta {\ensuremath{e}} states are separated from the Ta
{\ensuremath{t_{2}}} states by a band gap of 1.3 eV and occupy the energy
interval from 3.8 to 5.6 eV. Although formally the Ta valency in
Ta$_2$NiSe$_5$ is close to Ta$^{5+}$ (5$d^{0}$) the occupation number of 5$d$
electrons in the Ta atomic sphere is equal to 2.6. The excessive charge is
provided by the tails of Se 4$p$ and Ni 3$d$ states. These 5$d_{Se}$ and
5$d_{Ni}$ states play an essential role in the RIXS spectrum at the Ta $L_3$
edge (see Section V).

\section{X-ray absorption spectra}

\begin{figure}[tbp!]
\begin{center}
  \includegraphics[width=0.9\columnwidth]{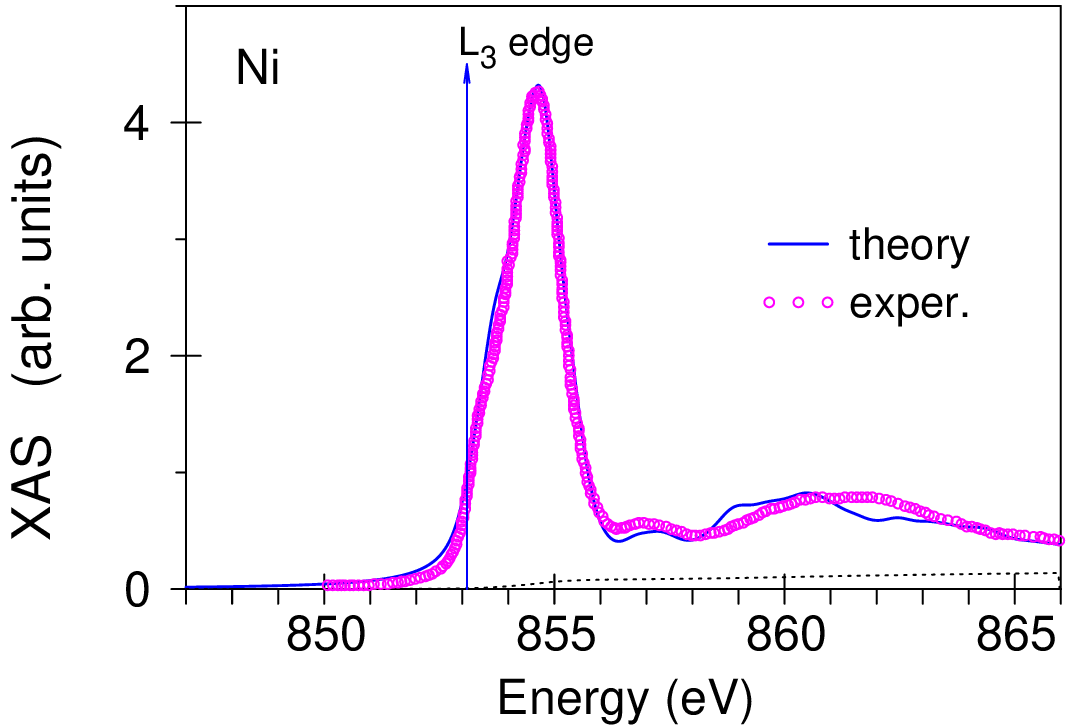}
\end{center}
\caption{\label{XAS_Ni_L3}(Color online) The theoretically calculated
  (the full blue curve) and experimentally measured \cite{MHP+20}
  (open magenta circles) x-ray absorption spectroscopy (XAS) spectra
  at the Ni $L_3$ edge. }
\end{figure}

\begin{figure}[tbp!]
\begin{center}
\includegraphics[width=0.9\columnwidth]{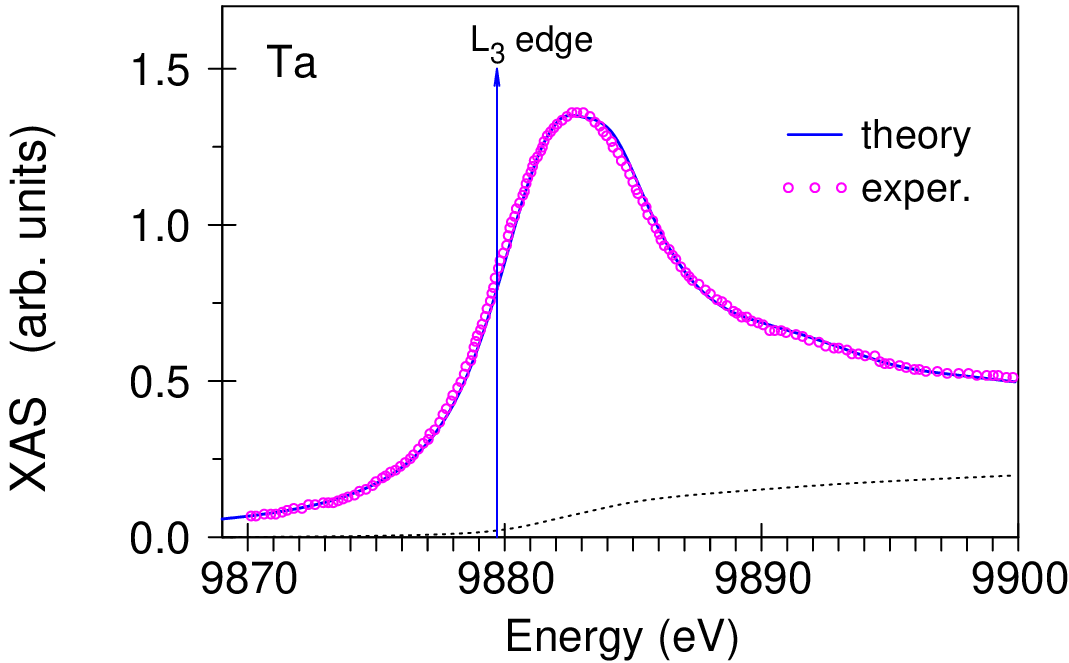}
\end{center}
\caption{\label{XAS_Ta_L3}(Color online) The theoretically calculated
  (the full blue curve) and experimentally measured \cite{LRK+21}
  (open magenta circles) x-ray absorption spectroscopy (XAS) spectra
  at the Ta $L_3$ edge. }
\end{figure}

Figures \ref{XAS_Ni_L3} and \ref{XAS_Ta_L3} show the theoretically
calculated (full blue curves) XAS spectra compared with the
experimentally measured spectra at the Ni $L_3$ \cite{MHP+20} and Ta
$L_3$ \cite{LRK+21} edges, respectively.  Although Ni $L_3$ XAS
extends more than 15 eV above the edge, the major peak situated at 1.5
eV above the edge is very narrow with a half-width of $\sim$2 eV. The
peak possesses a low energy shoulder at 0.6 eV above the edge. The
spectrum reflects the energy distribution of the Ni 3$d$ states which
are located between 0.16 and 2.5 eV (see Fig. \ref{PDOS_TNS}). The
major peak is created by the states situated between 1.2 and 2.5 eV,
and the low energy shoulder is due to the states at 0.16$-$1.0 eV. Ni
3$d$ PDOS is very small above 3.5 eV (see Fig. \ref{PDOS_TNS}), which
leads to relatively small fine structures in Ni $L_3$ XAS above 856
eV. The peak at 857 eV is due to transitions into the Ni 3$d$ empty
states which are derived from the hybridization with Ta 5$d$ states
while the fine structures between 859 and 866 eV are due to the
hybridization with free electron-like states.

Ta $L_3$ XAS has an almost four times larger half-width (Fig. \ref{XAS_Ta_L3})
in comparison with Ni $L_3$ XAS because it reflects the energy distribution of
the Ta {\ensuremath{t_{2}}} and {\ensuremath{e}} states situated at 0.16$-$2.5
eV and 3.8$-$5.6 eV, respectively. Besides, the width of the Ta 3$p_{3/2}$
core level is much larger than that of Ni 3$p_{3/2}$ (4.68 and 0.47 eV,
respectively \cite{CaPa01}). The band structure calculations reproduce well
the XAS spectra at the Ni and Ta $L_3$ edges.

\section{RIXS spectra}
\label{sec:rixs}

\subsection{N\lowercase{i} $L_3$ RIXS spectrum}
\label{sec:rixs_Ni}

The experimental RIXS spectrum at the Ni $L_3$ edge in Ta$_2$NiSe$_5$
was measured by Monney {\it et al.} \cite{MHP+20} and Lu {\it et al.}
\cite{LRK+21} in the energy interval up to 6 and 1.2 eV,
respectively. The RIXS spectrum occupies a relatively small energy
interval and possesses several fine structures.

\begin{figure}[tbp!]
\begin{center}
\includegraphics[width=0.9\columnwidth]{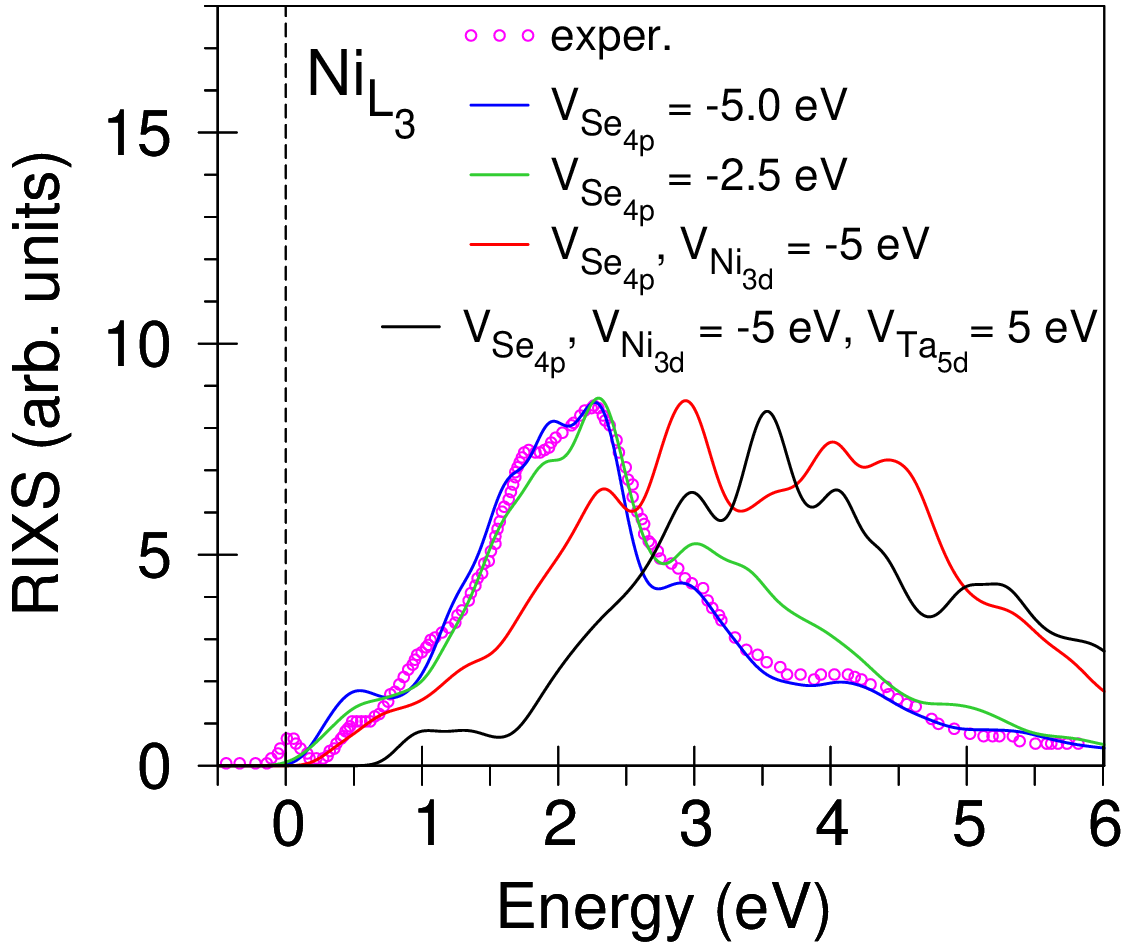}
\end{center}
\caption{\label{rixs_Ni_Vl}(Color online) The experimental resonant
  inelastic x-ray scattering (RIXS) spectrum at the Ni $L_3$ edge
  \cite{MHP+20} in near-specular geometry {\bf Q} = (0.0625, 1.25, 0)
  in reciprocal lattice units for incident photon energy
  $\hbar \omega_{in}$ = 853.3 eV compared with the theoretical RIXS
  spectra for the same geometry and the incident photon energy
  calculated in the GGA+SIC+SO approximation for different parameters
  $V_l$. }
\end{figure}

Figure \ref{rixs_Ni_Vl} presents the influence of the SIC parameter
$V_l$ on the shape of the Ni $L_3$ RIXS spectrum. We obtained
reasonably good agreement with the experiment applying $V_l$ only to
the Se 4$p$ states. Using $V_{Se_{4p}}$ = $-$2.5 eV, we describe
better the low-energy part of the spectrum, while its high-energy part
is described better by $V_{Se_{4p}}$ = $-$5.0 eV.  Applying the SIC
parameter to Ni 3$d$ and/or Ta 5$d$ states makes the agreement with
the experiment only worser.

\begin{figure}[tbp!]
\begin{center}
\includegraphics[width=0.9\columnwidth]{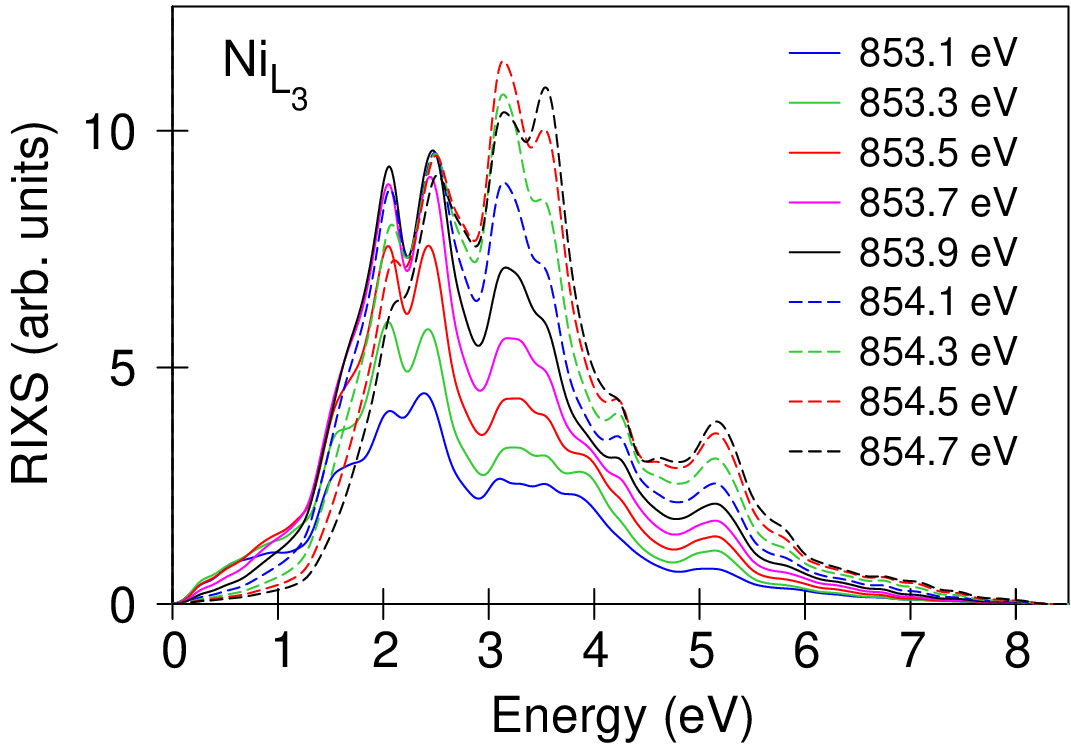}
\end{center}
\caption{\label{rixs_Ni_Ei}(Color online) Resonant inelastic x-ray scattering
  (RIXS) spectra as a function of incident photon energy, calculated at the Ni
  $L_3$ edge with momentum transfer vector {\bf Q} = (0.0625, 1.25, 0) in
  reciprocal lattice units. }
\end{figure}

Figure \ref{rixs_Ni_Ei} shows the Ni $L_3$ RIXS spectrum as a function
of incident photon energy above the corresponding edge. We found that
the fine structure between 3 and 4 eV is increased with increasing the
incident photon energy. It is in qualitative agreement with the
corresponding experimental dependence, however, the experiment shows
weaker dependence \cite{MHP+20}.

\begin{figure}[tbp!]
\begin{center}
\includegraphics[width=0.9\columnwidth]{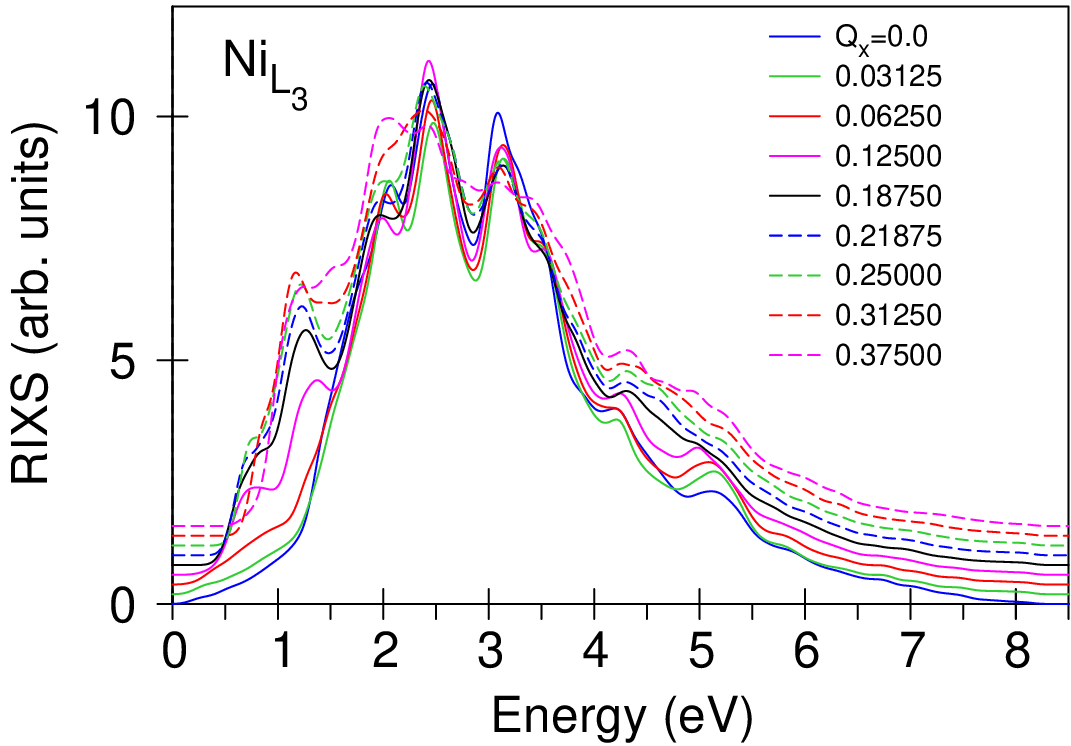}
\end{center}
\caption{\label{rixs_Ni_Qx}(Color online) Resonant inelastic x-ray scattering
  (RIXS) spectra at the Ni $L_3$ edge calculated as a function of $Q_x$ in the
  momentum transfer vector {\bf Q} = (Q$_x$, 1.25, 0) for incident photon
  energy $\hbar \omega_{in}$ = 853.3 eV. }
\end{figure}

It is widely believed that the $d-d$ excitations show only small
momentum transfer vector {\bf Q} dependence in 5$d$ transition metal
compounds \cite{LKH+12,KTD+20}. However, the soft RIXS spectra in 3$d$
transition metals are more sensitive to the value of {\bf Q}. Figure
\ref{rixs_Ni_Qx} shows RIXS spectra at the Ni $L_3$ edge calculated as
a function of $Q_x$ in {\bf Q} = (Q$_x$, 1.25, 0) for incident photon
energy $\hbar \omega_{in}$ = 853.3~eV. We found that with the
increasing of Q$_x$ the low-energy peak at 1 eV is significantly
increased. The fine structure at 2 eV is also increased but to a
lesser extent. The similar dependence was also observed
experimentally, however, it is weaker in the experiment \cite{MHP+20}.

\begin{figure}[tbp!]
\begin{center}
\includegraphics[width=0.9\columnwidth]{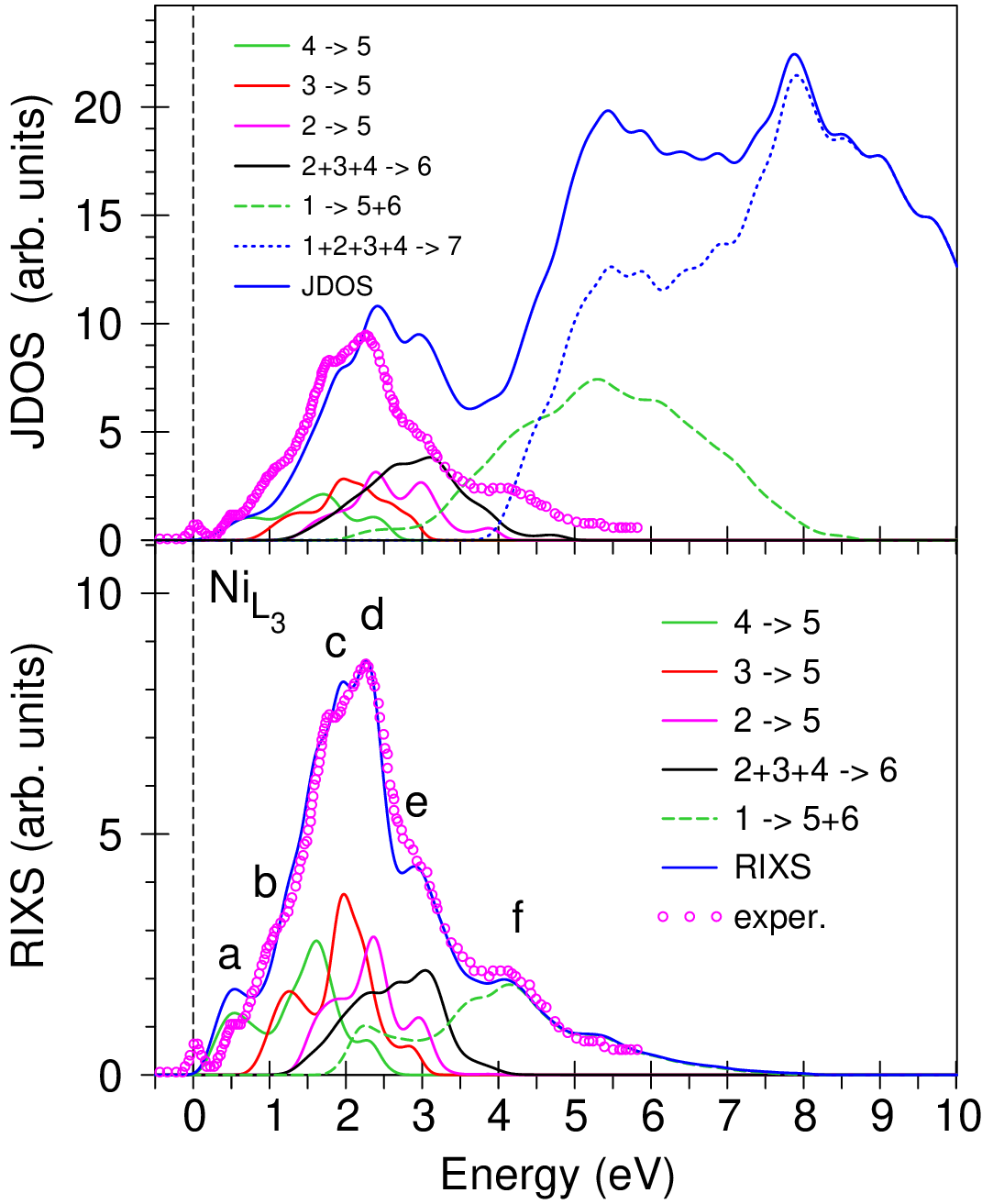}
\end{center}
\caption{\label{rixs_Ni_JDOS}(Color online) The lower panel: the
  experimentally measured resonant inelastic x-ray scattering (RIXS)
  spectrum at the Ni $L_3$ edge \cite{MHP+20} (open magenta circles)
  in near-specular geometry {\bf Q} = (0.0625, 1.25, 0) for incident
  photon energy $\hbar \omega_{in}$ = 853.3 eV compared with the
  theoretical RIXS spectra calculated for the same geometry and
  incident photon energy in the GGA+SIC+SO approximation with
  $V_{Se_{4p}}$ = $-$5.0 eV (the full blue curve) and partial
  contributions from different interband transitions presented in
  Fig. \ref{BND_DOS_Ni}. The upper panel: joint density of states
  (JDOS) (the full blue curve) and partial transitions from different
  interband transitions in comparison with the experimental RIXS
  spectrum (open magenta circles).}
\end{figure}

\begin{figure}[tbp!]
\begin{center}
\includegraphics[width=1.\columnwidth]{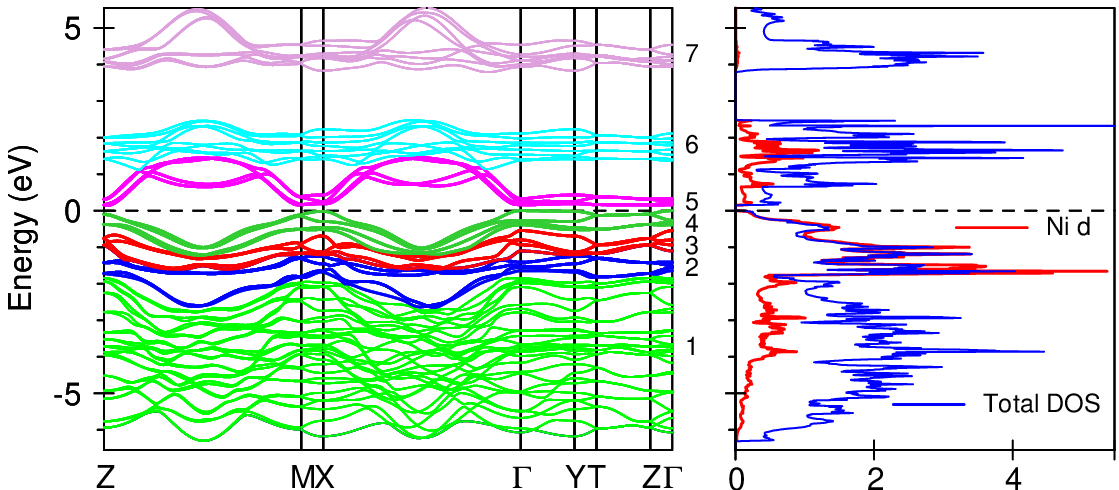}
\end{center}
\caption{\label{BND_DOS_Ni}(Color online) The energy band structure, partial
  Ni 3$d$ density of states (DOS) [in states/(atom eV)] (the red curve) and
  normalized total DOS (the blue curve) of monoclinic Ta$_2$NiSe$_5$
  calculated in the fully relativistic Dirac GGA+SIC+SO approximation
  ($V_{Se_{4p}}$ = $-$5.0 eV). }
\end{figure}

The experimentally measured RIXS spectrum consists of a peak centered
at zero energy loss, which comprises the elastic line and other
low-energy features such as phonons, magnons, etc., and at least six
inelastic excitations. Figure \ref{rixs_Ni_JDOS} (the lower panel)
presents the experimental RIXS spectrum at the Ni $L_3$ edge
\cite{MHP+20} (open magenta circles) for {\bf Q} = (0.0625, 1.25, 0)
in reciprocal lattice units and incident photon energy
$\hbar \omega_{in}$ = 853.3 eV compared with the theoretical RIXS
spectrum calculated for the same geometry in the GGA+SIC+SO
approximation with $V_{Se_{4p}}$ = $-$5.0 eV (the full blue curve) and
partial contributions from different interband transitions presented
in Fig. \ref{BND_DOS_Ni} by different colors. We can divide the Ni
3$d$ valence band in Fig. \ref{BND_DOS_Ni} into two groups. The bands
number 1 situated between $-$6.3 and $-$1.8 eV are the Ni
3$d_{\rm{Se}}$ states which are derived from the decomposition of Se
4$p$ states inside the Ni atomic spheres. The bands situated between
$-$1.8 eV and $E_F$ are Ni 3$d$ states themselves. We divide the
latter bands into three groups with numbers from 2 to 4. The empty Ni
3$d$ bands we subdivide into two groups with numbers 5 and 6. We found
that the first four fine structures ($a$, $b$, $c$, and $d$) of the Ni
$L_3$ RIXS spectrum are derived from interband transitions from the
occupied bands with numbers 2 to 4 to the empty bands number 5. The
low-energy peak $a$ is due to interband transition between the closest
to the Fermi level bands 4 and 5. The same transitions also contribute
to the low-energy shoulder $c$. The major peak $d$ is derived from 3
$\rightarrow$ 5 and 4 $\rightarrow$ 5 transitions. The high-energy
fine structure $d$ is due to interband transitions from all occupied
bands (besides bands 1) to the group of empty bands by number 6. The
highest energy structure $f$ is due to transitions from the Ni
3$d_{\rm{Se}}$ states (bands 1) to empty states with numbers 5 and
6. It is important to note that all these interband transitions
strongly overlap with each other.

\begin{figure}[tbp!]
\begin{center}
\includegraphics[width=1.\columnwidth]{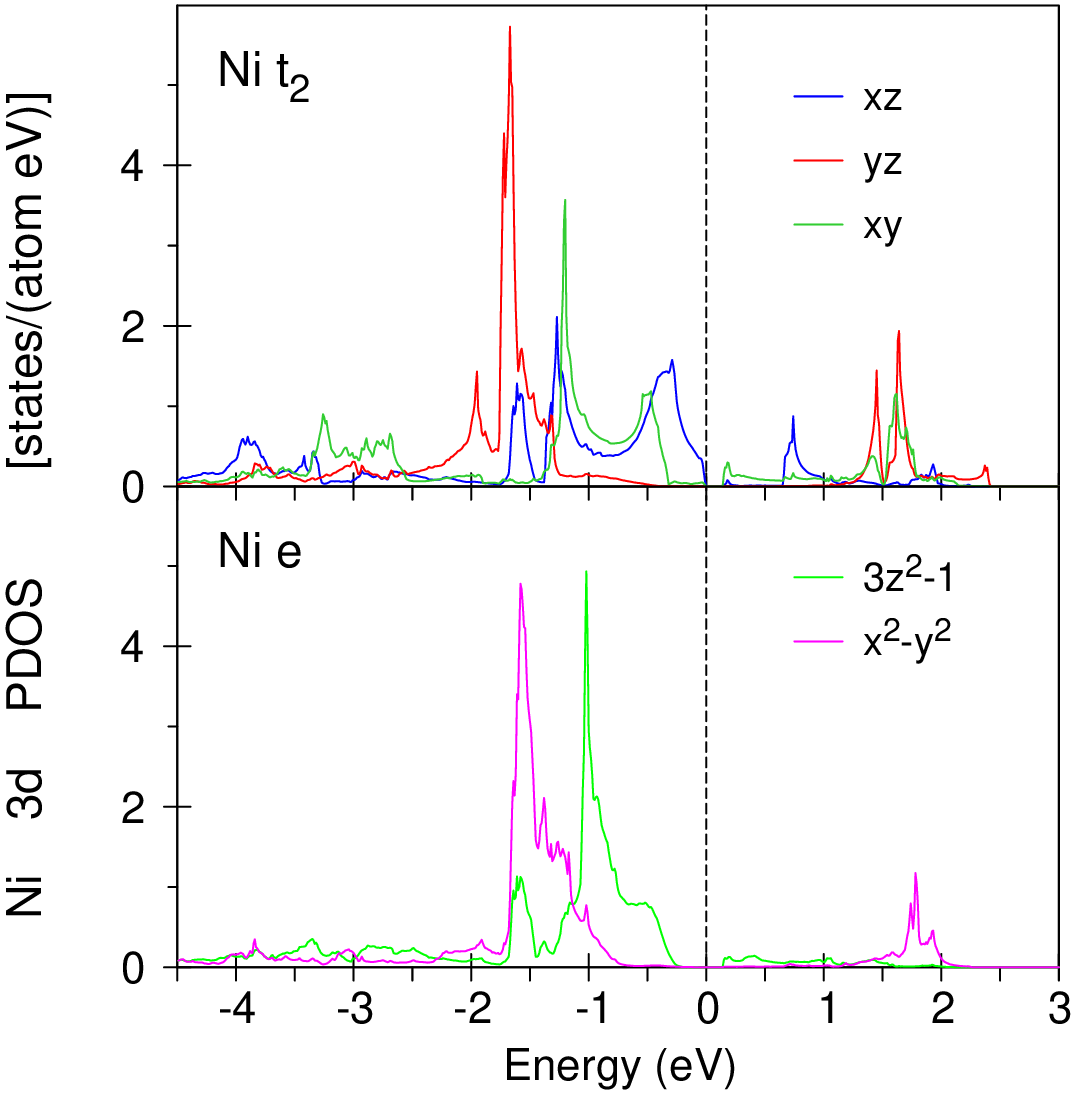}
\end{center}
\caption{\label{PDOS_Ni_orbit}(Color online) Orbital resolved Ni 3$d$ density
  of states (DOS) [in states/(atom eV)] of monoclinic Ta$_2$NiSe$_5$
  calculated in the fully relativistic Dirac GGA+SIC+SO approximation
  ($V_{Se_{4p}}$ = $-$5.0 eV). }
\end{figure}

Analyzing different interband transitions we found that
(2+3+4)$\rightarrow$5 transitions are larger than
(2+3+4)$\rightarrow$6 transitions. RIXS is an element- and
orbital-selective X-ray spectroscopy technique, based on a two-step,
two-photon resonant process. It combines X-ray emission spectroscopy
(XES) with X-ray absorption spectroscopy by measuring the coherent
X-ray emission at an incident X-ray photon energy within the near edge
X-ray absorption spectrum. In the first step (X-ray absorption) in our
case, an electron of the absorbing atom is resonantly excited from a
2$p_{3/2}$ core level into a 5 or 6 empty energy bands. In the second
step (X-ray emission), the system radiatively decays from the (2+3+4)
occupied bands into the 2$p_{3/2}$ core level, accompanied by a
photon-out emission. Because the X-ray emission is the same for both
the cases, we have consider only absorption process into bands 5 and 6
separately. Ref. \cite{AYJ10} presents the angular matrix elements for
dipole allowed transitions at the $L_3$ edge from initial core states
with different projections $m_j$ of the total angular momentum $j=3/2$
to $d$ cubic harmonics. It was shown that for the $\sigma$ incident
light the largest contribution is due to 2$p_{3/2}$ $\rightarrow$
{\dxy} and 2$p_{3/2}$ $\rightarrow$ {\dxxyy} transitions (two times
larger than the transitions into {\dyz} and {\dxz} states and three
times larger than the transitions into {\dzz} states). Figure
\ref{PDOS_Ni_orbit} presents orbital resolved Ni 3$d$ DOS. The bands
number 5 are derived mostly from {\dxy} orbitals, bands 6 are from
{\dxz}, {\dyz}, and {\dzz} orbitals.  It explains why the
(2+3+4)$\rightarrow$5 transitions exceed the (2+3+4)$\rightarrow$6
ones.

The upper panel of Fig. \ref{rixs_Ni_JDOS} shows JDOS (the full blue
curve) and partial transitions from different interband transitions in
comparison with the experimental RIXS spectrum. JDOS is not able to
describe correctly the the RIXS spectrum above 2 eV. It can be
explained by the difference in the shape of the total and partial Ni
3$d$ DOS presented in Fig. \ref{BND_DOS_Ni} (blue and red curves on
the right panel, respectively). We normalized total DOS to Ni PDOS at
$-$1.6 eV. Total DOS of the band group number 1 is significantly
larger in comparison with Ni PDOS at the same energy interval. As a
result, the intensity of 1 $\rightarrow$ (5+6) transitions is
significantly increased for JDOS. Besides, new very intensive
transitions (1+2+3+4) $\rightarrow$ 7 appear, which are completely
absent in the theoretically calculated RIXS spectrum. The RIXS
spectrum at the Ni $L_3$ edge can be correctly described only with
taking into account corresponding matrix elements.

\subsection{T\lowercase{a} $L_3$ RIXS spectrum}
\label{sec:rixs_Ta}

\begin{figure}[tbp!]
\begin{center}
\includegraphics[width=0.9\columnwidth]{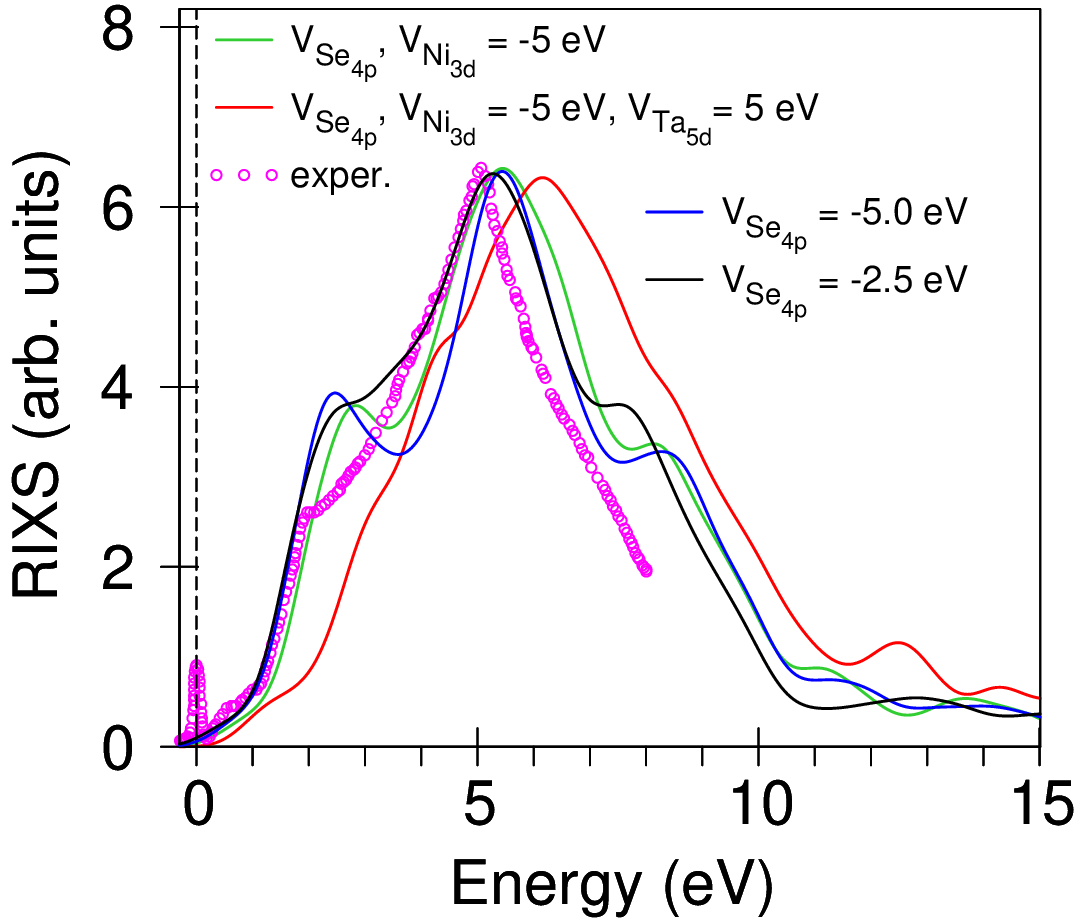}
\end{center}
\caption{\label{rixs_Ta_Vl}(Color online) The experimental resonant inelastic
  x-ray scattering (RIXS) spectrum at the Ta $L_3$ edge compared with the
  theoretical RIXS spectra calculated in the GGA+SIC+SO approximation for
  different parameters $V_l$. }
\end{figure}

Figure \ref{rixs_Ta_Vl} presents the influence of the SIC parameter
$V_l$ on the shape of the Ta $L_3$ RIXS spectrum. This parameter is
less critical than in the Ni $L_3$ RIXS spectrum. The combination of
the parameters ($V_{Ta_{5d}}$, $V_{Ni_{3d}}$, $V_{Se_{4p}}$) = (0,
$-$5, $-$5), (0, 0, $-$5), and (0, 0, $-$2.5) eV give similar Ta $L_3$
RIXS spectra, although, the last combination produces a slightly
better description of the low-energy shoulder at $\sim$2.5
eV. Applying an additional SIC parameter $V_l$ to the Ta 5$d$ states
makes the agreement with the experiment worse.

\begin{figure}[tbp!]
\begin{center}
\includegraphics[width=0.9\columnwidth]{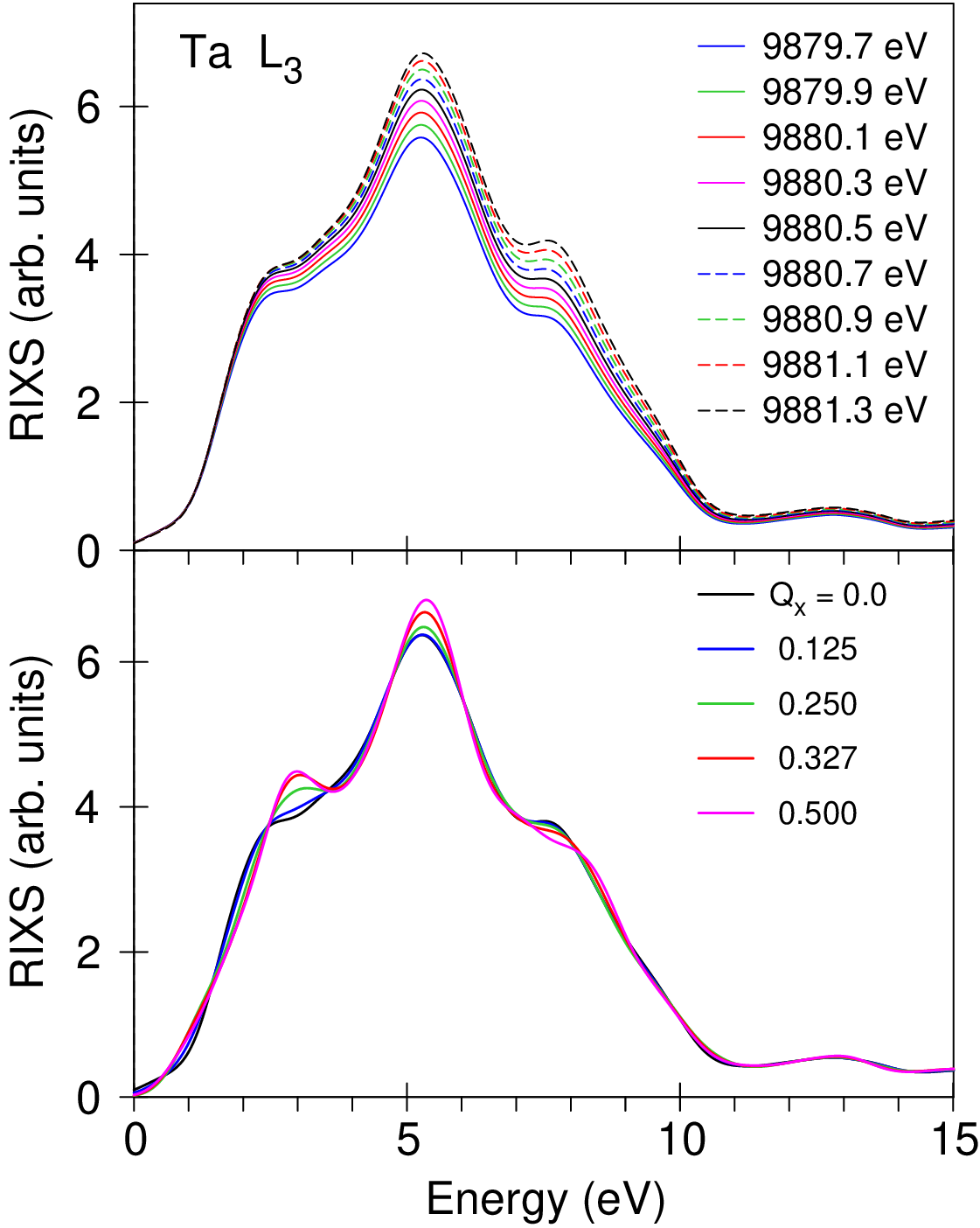}
\end{center}
\caption{\label{rixs_Ta_Ei_Qx}(Color online) The lower panel: resonant
  inelastic x-ray scattering (RIXS) spectra at the Ta $L_3$ edge
  calculated as a function of $Q_x$ in the momentum transfer vector
  {\bf Q} = (Q$_x$, 14.5, 0) for incident photon energy
  $\hbar \omega_{in}$ = 9679.7 eV. The upper panel: RIXS spectra as a
  function of incident photon energy calculated at the Ta $L_3$ edge
  with {\bf Q} = (0, 14.5, 0) in reciprocal lattice units. }
\end{figure}

Figure \ref{rixs_Ta_Ei_Qx} shows the theoretically calculated RIXS
spectra at the Ta $L_3$ edge as a function of incident photon energy
(the upper panel), and as a function of $Q_x$ in the momentum transfer
vector {\bf Q} = (Q$_x$, 14.5, 0) (the lower panel). RIXS spectrum at
the Ta $L_3$ edge possesses relatively weak dependence from incident
photon energy as well as from the momentum transfer vector {\bf Q} in
comparison with the corresponding dependences at the Ni $L_3$ edge
(compare Fig. \ref{rixs_Ta_Ei_Qx} with Figs. \ref{rixs_Ni_Ei} and
\ref{rixs_Ni_Qx}). It is in agreement with the Krajewska {\it et al.}
who claimed that the $d-d$ excitations show only a small momentum
transfer vector {\bf Q} dependence in the 5$d$ transition metal
compounds \cite{KTD+20}.

\begin{figure}[tbp!]
\begin{center}
\includegraphics[width=0.9\columnwidth]{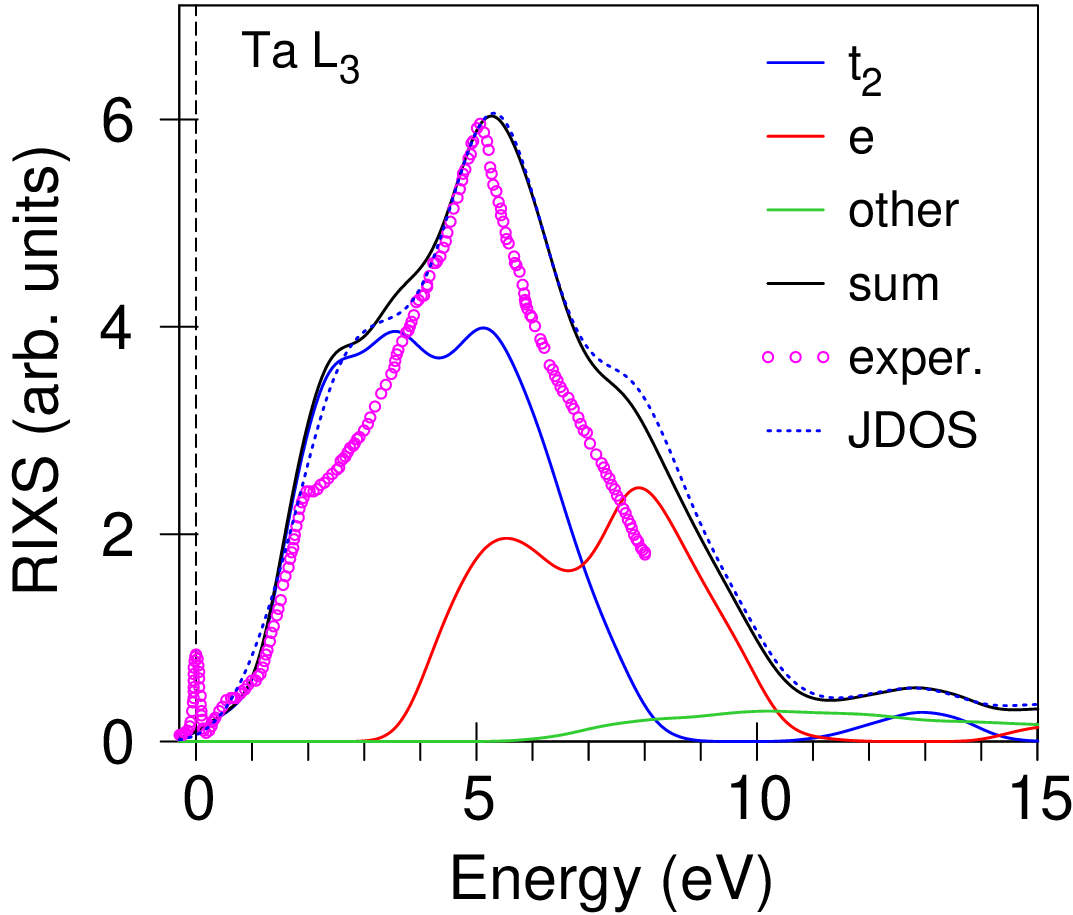}
\end{center}
\caption{\label{rixs_Ta_L3}(Color online) The experimental resonant
  inelastic x-ray scattering (RIXS) spectrum at the Ta $L_3$ edge
  (open magenta circles) compared with the theoretically calculated
  one (the full black curve) and partial contributions from different
  interband transitions for the momentum transfer vector {\bf Q} = (0,
  14.5, 0) in reciprocal lattice units. The dotted blue curve presents
  joint density of states (JDOS). }
\end{figure}

Figure \ref{rixs_Ta_L3} presents the experimental RIXS spectrum at the Ta
$L_3$ edge compared with the theoretically calculated one (the full black
curve) and partial contributions from different interband transitions. Because
the Ta valency in Ta$_2$NiSe$_5$ is close to Ta$^{5+}$ (5$d^{0}$) the Ta $L_3$
RIXS spectrum is formed by a charge transfer between occupied 5$d_{Se}$ and
5$d_{Ni}$ states (see Fig. \ref{PDOS_TNS}) and empty {\ensuremath{t_{2}}} and
{\ensuremath{e}} states (blue and red curves in Fig. \ref{rixs_Ta_L3},
respectively). The transitions into the other empty states are very small (the
green curve in Fig. \ref{rixs_Ta_L3}).  Empty total DOS in Ta$_2$NiSe$_5$ is
formed mostly by Ta 5$d$ states, besides, the energy distribution of the
occupied part of Ta 5$d$ PDOS, which is formed by the tails of Se 4$p$ and Ni
3$d$ states inside the Ta atomic spheres, repeats the shape of total DOS,
therefore, JDOS describes quite well the RIXS spectrum at the Ta $L_3$ edge
(the dotted blue curve in Fig. \ref{rixs_Ta_L3}).

\section{Conclusions}

The electronic and magnetic properties of quasi-one-dimensional
Ta$_2$NiSe$_5$ were investigated theoretically in the frame of the
fully relativistic spin-polarized Dirac LMTO band-structure method.
We found that SOC plays a minor role in the electronic structure of
Ta$_2$NiSe$_5$.  The GGA approximation produces a metallic ground
state in Ta$_2$NiSe$_5$ in contradiction with ARPES, dc transport, and
optical measurements, which indicate that Ta$_2$NiSe$_5$ is a small
band-gap semiconductor. To obtain the semiconducting ground state in
Ta$_2$NiSe$_5$ we use a SIC-like orbital-dependent potential $V_l$
incorporated into the Hamiltonian. Although the correct value of the
band gap can be achieved using different combinations of the
parameters $V_{Ta_{5d}}$, $V_{Ni_{3d}}$, and $V_{Se_{4p}}$, the shift
of Se 4$p$ states is most important in Ta$_2$NiSe$_5$. To reproduce
the experimental energy gap and get the best agreement between the
theory and different experiments we applied the orbital-dependent
potential $V_l$ for the Se 4$p$ states only. We found that the value
of $V_{Se_{4p}}$ is somewhere between $-$2.5 and $-$5.0 eV. The value
$V_{Se_{4p}}$ = $-$2.5 eV describes better the low-energy parts of Ni
and Ta $L_3$ RIXS spectra, while the value $-$5.0 eV describes better
the high-energy part of the Ni RIXS spectrum and produces the correct
value of the energy band gap.

We investigated experimentally the RIXS spectrum at the Ta $L_3$ edge
and theoretically at the Ni and Ta $L_3$ edges in Ta$_2$NiSe$_5$. The
experimentally measured RIXS spectrum at the Ni $L_3$ edge in addition
to the elastic scattering peak at 0 eV possesses several features. We
interpret these structures by analyzing particular interband
transitions. We investigated the RIXS spectra at the Ni and Ta $L_3$
edges as a function of momentum transfer vector {\bf Q} and incident
photon energy. The RIXS spectrum at the Ta $L_3$ edge possesses
relatively weak dependence on the incident photon energy as well as on
the {\bf Q} vector. The RIXS spectrum at the Ni $L_3$ edge shows a
strong increase of the low-energy peak at $\sim$1 eV with an increase
of the Q$_x$ component of the vector {\bf Q}. The increase of the
incident photon energy $\hbar\omega_{in}$ above the Ni $L_3$ edge
leads to an increase of the RIXS spectrum in the energy range from 3
to 4 eV.

Empty total DOS in Ta$_2$NiSe$_5$ is formed mostly by Ta 5$d$ states
and the energy distribution of the occupied part of Ta 5$d$ PDOS,
which is formed by the tails of the Se 4$p$ and Ni 3$d$ states inside
the Ta atomic spheres, repeats the shape of total DOS, therefore, JDOS
describes quite well the RIXS spectrum at the Ta $L_3$ edge. However,
the Ni $L_3$ RIXS spectrum can be correctly described only with taking
into account corresponding matrix elements.

\section*{Acknowledgments}

RIXS experiments at the Ta $L_3$-edge were performed at the BL11XU of
SPring-8 with the approval of the Japan Synchrotron Radiation Research
Institute (JASRI) (Proposals No. 2016A3552 and No. 2016BA3552).


\newcommand{\noopsort}[1]{} \newcommand{\printfirst}[2]{#1}
  \newcommand{\singleletter}[1]{#1} \newcommand{\switchargs}[2]{#2#1}

\end{document}